\documentclass[preprint2]{emulateapj}

\slugcomment{version 2010 Feb 2: fm}

\shorttitle{Fueling lobes in radio galaxies}
\shortauthors{Massaro F. \& Ajello, M.}

\newcommand{\chn}{{\it Chandra}}
\newcommand{\xmm}{{\it XMM-Newton}}
\newcommand{\fer}{{\it Fermi}}
\newcommand{\suz}{{\it Suzaku}}

\begin{document}

\title{Fueling lobes of radio galaxies:\\statistical particle acceleration and the extragalactic $\gamma$-ray background}%@title

\author{F. Massaro\altaffilmark{1} and M. Ajello\altaffilmark{2}}
\affil{Harvard, Smithsonian Astrophysical Observatory, 60 Garden Street, Cambridge, MA 02138, USA}
\affil{SLAC National Laboratory and Kavli Institute for Particle Astrophysics and Cosmology, 2575 Sand Hill Road, Menlo Park, CA 94025, USA}

\begin{abstract} 
The recent discovery of the $\gamma$-ray emission from the lobes of the closest radio galaxy Centaurus A by \fer~implies the presence 
of high-energy electrons at least up to $\gamma \sim 10^5$ - 10$^6$. 
These high-energy electrons are required to interpret the observed $\gamma$-ray radiation in terms of 
inverse Compton emission off the cosmic microwave background (IC/CMB);
the widely accepted scenario to describe the X-ray emission of radio galaxy lobes.
In this Letter, we consider the giant radio lobes of FR II radio galaxies 
showing that it is possible to maintain electrons at energies $\gamma$ $\sim$ 10$^5$ - 10$^6$,
assuming an acceleration scenario (driven by turbulent magnetic fields) that compensates the radiative losses.
In addition, we consider the contribution to the diffuse extragalactic $\gamma$-ray background due to the IC/CMB emission of  FR\,IIs' lobes 
showing its relevance in the keV to MeV energy range. 
\end{abstract}


\section{Introduction}
Radio galaxies exhibit several extended  components on the kpc scale, namely: jets, hotspots and lobes.
The lobes are double structures extended over large scales,
roughly symmetrical and ellipsoidal lying on both sides of the active nuclei of radio galaxies.

It was first noticed by Fanaroff \& Riley (1974) that the relative positions of regions of 
high and low surface brightness in the extended components of extragalactic radio sources are 
correlated with their radio luminosity. 
It was found that nearly all sources with luminosity
$L_{178MHz}$ $\leq$  2 $\times$ 10$^{25}$ $h_{100}^{-2}$ W Hz$^{-1}$ str$^{-1}$
were of Class I (i.e., FR\,I) while the brighter sources were nearly all of Class II (i.e., FR\,II). 
FR\,Is have a surface brightness which is larger toward their cores while FR\,IIs have it larger toward their edges.

The morphology of radio galaxies turns out to reflect the method 
of energy transport in the radio source. FR\,I radio galaxies typically have bright jets in the center, 
while FR\,IIs have faint jets but bright hotspots at the ends of their lobes (e.g., Baum et al. 1995). 
 
The radio emission arising from the lobes is widely interpret as due 
to synchrotron radiation,
while the high-energy radiation has been successfully described 
as inverse Compton emission off the cosmic microwave background (IC/CMB).
Moreover,
because of the low electron densities in the large lobe volumes 
the synchrotron self Compton (SSC) mechanism produces a negligible contribution
both at X-ray and at $\gamma$-rays (e.g., Hardcastle et al. 2002, Croston et al. 2005, hereafter C05).

The radio lobes extension is typically $\sim$ 100 kpc
implying a timescale for their growth spanning from tens to hundreds of millions of years (Blundell \& Rawlings 2000, Mullin et al. 2008).
If relativistic particles in the lobes are not continuously accelerated, 
considering the synchrotron cooling time $\tau_{syn}$ $\sim$ 16.4 $\cdot$ $\gamma^{-1}$ $B^{-2}$(G) yr,
relativistic electrons with Lorentz factor $\gamma$ $\geq$ 10$^6$ in magnetic fields of $\sim$ 10$^{-5}$ - 10$^{-6}$ G
do not emit on epochs longer than $\sim$ 1 Myr, which is shorter than the age of radio lobes
(van der Laan \& Perola 1969).

A population of energetic electrons is required to interpret the radio/millimeter observations of 
radio galaxy lobes (e.g., Hardcastle \& Looney 2008), and recently, the {\it Wilkinson Microwave Anisotropy Probe} (WMAP) and \fer~observations 
of the Centaurus A lobes (Hardcastle et al. 2009: Abdo et al. 2010a).
In addition, the \chn, \xmm~ and \suz~observations of radio galaxy lobes (Isobe et al. 2002, 2005, 2009, Kataoka et al. 2003, C05) 
show that their IC/CMB spectral energy distribution (SED) is rising in the X-rays
and consequently a significant fraction of their high-energy emission is expected to be radiated in the hard X-rays and at low energy $\gamma$-rays.

Following the idea of Bergamini et al. (1967), the IC/CMB process of extended structures in radio galaxies
could provide a significant contribution to the diffuse extragalactic $\gamma$ ray background (EGB) in the keV to MeV energy range.

The contribution to the EGB of kpc-scale jets in the case of FR\,Is, from IC scattering of starlight photons by the ultrarelativistic
electrons, has been estimated by Stawarz et al. (2006). These estimates show that this emission contributes 
$\sim$ 1\% of the EGB at GeV energies, and it is less relevant at MeV energies, where the contribution of the IC/CMB emission peaks. 

In this Letter, we adopt a statistical acceleration scenario for the lobes in FR\,IIs, due to turbulent magnetic fields, that balances the radiative cooling and 
keeps electron energies up to $\gamma$ $\sim$ 10$^5$ - 10$^6$ (e.g., Pacholczyk \& Scott 1976, Eilek \& Shore 1989),
in agreement with the recent \fer~observations.
A similar mechanism has been proposed by Hardcastle et al. (2009) and O'Sullivan et al. (2009)
to identify radio galaxy lobes as possible accelerators for the ultra high-energy cosmic rays (e.g., Hillas 1984).

Finally, we consider a template SED of the emission from the FR\,IIs' lobes to estimate their contribution to the EGB.

We use cgs units and a flat cosmology with $H_0=72$ km s$^{-1}$ Mpc$^{-1}$,
$\Omega_{M}=0.26$ and $\Omega_{\Lambda}=0.74$ (Dunkley et al. 2009).

\section{Particle energy losses in radio galaxy lobes}
The relativistic electrons in radio galaxy lobes are subject to two main radiative loss processes:
the synchrotron radiation and the IC/CMB. The first mechanism is responsible for the lobe radio emission, while
the latter describes their observed high-energy emission in the X-rays, and recently, also in the $\gamma$-rays
as in the nearby case of Centaurus A (Abdo et al. 2010a).

To clarify these radiative processes it is interesting to consider the case of a 
single relativistic electron of Lorentz factor $\gamma$.
The total power emitted via synchrotron and IC/CMB radiation is:
\begin{equation}
P_{rad}= 1.05 \cdot 10^{-15}\,\gamma^2\, (B^2+B_{CMB}^2)\, \rm{erg\,s^{-1}},
\end{equation}
where $B$ is the lobe magnetic field while
$B_{CMB}$ = 3.26 $\cdot$ 10$^{-6}$\,$(1+z)^2$ G, the value of the magnetic field that has the same energy density as the 
CMB at redshift $z$ (e.g., Murgia et al. 1999).
The $B/B_{CMB}$ ratio identifies which emission process, synchrotron or IC/CMB,
is more relevant for the electron radiative cooling.
%-----------------------------------------------------------------------------------------------------------------------
\begin{figure}[!htp]
\includegraphics[height=7.5cm,width=8.8cm,angle=0]{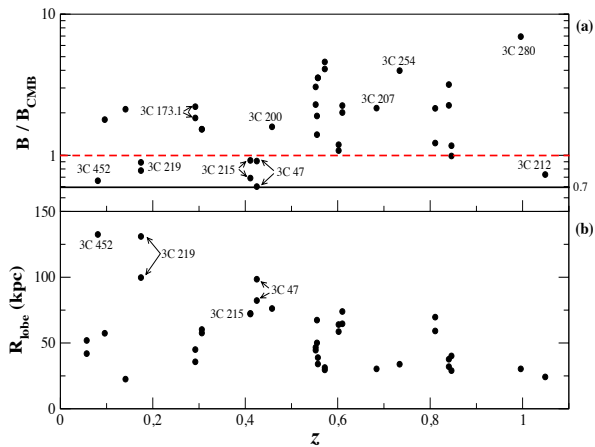}
\caption{a) The $B/B_{CMB}$ ratio (upper panel) 
for the FR\,IIs lobes in the sample of C05. The dashed red line indicates the case $B/B_{CMB}$ = 1, 
when synchrotron and IC/CMB losses have the same relevance.
b) The equivalent spherical radius of the lobe volume, $R_{lobe}$, as a function of 
redshift estimated for the same sample.}
\end{figure}
%-----------------------------------------------------------------------------------------------------------------------
In Figure 1(a), we report the $B/B_{CMB}$ ratio and $R_{lobe}$, the equivalent spherical radius of the lobe volume, with respect to $z$,
for the sample of FR\,II lobes detected in the X-rays (C05), where
$B$ has been estimated by fitting the radio-to-X-ray lobe SEDs.

We note that the ratio $B/B_{CMB}$ is of order a few, in particular for the sources with larger $R_{lobe}$ at low redshift (i.e. $z \leq$0.5), 
for which the CMB energy density exceeds the estimate of the magnetic field energy density (Figure 1(b)).
Seventy-five percent of the considered FR\,II lobes have $B/B_{CMB}$ in the range 0.7--2 (Figure 1a), 
suggesting that in these sources the radiative losses for synchrotron and for IC/CMB have similar importance.
This strongly suggests that the luminosity ratio of the two SED components $L_{syn}/L_{ic}$ is $\sim$ 1.

However, these estimates of the $B/B_{CMB}$ ratio for radio galaxy lobes are dependent 
on the choice of the particle energy distribution (PED) adopted.
The restricted energy ranges, where lobe emission is detected, do not allow us
to estimate the PED shape and to constrain the $L_{syn}/L_{ic}$ accurately, introducing 
uncertainties in the estimates of the $B/B_{CMB}$.

Another relevant process for particle energy losses in lobes is adiabatic expansion. 
These losses could play a significant role during the evolution of radio lobes (e.g., Longair et al. 1973). 
Assuming a self similar expansion scenario (e.g., Matthews \& Scheuer 1990, Kaiser \& Alexander 1997),
the adiabatic losses are relevant in the initial phase of the lobe evolution, as in the case of young, compact radio galaxies. 

During the final phase of the lobe expansion,
when the lobe pressure is in equilibrium with that of the external medium,
losses are dominated by radiative processes, e.g., via IC/CMB or synchrotron emission (if $B/B_{CMB}$ is still $\gg$ 1).
FR\,IIs, with lobe size $\geq$ 30 - 50 kpc, can be considered 
at the end of their evolution, when the adiabatic expansion losses are negligible (e.g., Pacholczyk \& Scott 1976).

\section{Statistical acceleration scenario}
Acceleration by turbulent magnetic fields (due to electron scattering with magnetized plasma irregularities) 
via second order \fer~mechanisms, could be considered as a possible mechanism to
compensate both the synchrotron and the IC radiative losses (e.g., Eilek \& Shore 1989,
O'Sullivan et al. 2009, Hardcastle 2010).

Adopting this scenario, the acceleration timescale via statistical mechanism can be written as $\tau_{acc}$ $\simeq$
$l/(c\beta_A^2)$, where $l$ is the characteristic length scale of the magnetic field inhomogeneities (i.e., the mean free path of the electrons
between collisions) and $\beta_A$ = $u_A/c$ is the ratio of the Alfv\'en velocity $u_A$ and the speed of light $c$.
Consequently, according to Kardashev (1962), the energy gain
by systematic statistical acceleration, for a single relativistic electron in a turbulent magnetic field $B$ is:
\begin{equation}
\left(\frac{dE}{dt}\right)_{acc} =  \frac{u_A^2}{l\,c}\,E = \frac{m_e~c~u_A^2}{l}\,\gamma, 
\end{equation}
where $\rho_p~u_A^2/2 \sim B^2/8\pi$ is the energy density of the turbulent magnetic field, $n \simeq \rho_p/m_p$ 
is the plasma density and $m_e$ and $m_p$ the electron and the proton mass, 
respectively (e.g., Tsytovich 1966, Eilek \& Shore 1989).

We can estimate the maximum Lorentz factor of the electrons in the lobes, balancing 
the acceleration energy gain (Equation (2)) with the radiative losses (Equation (1)):
\begin{equation}
\gamma_{max} =  1.24 \times 10^{21} \left(\frac{1}{n\,l}\right) \left(\frac{B^2}{B^2+B^2_{CMB}}\right),
\end{equation} 
that is only dependent on the length scale, $l$, on the plasma density, $n$, and on the $B/B_{CMB}$ ratio.

Lobes are usually assumed to be close to the equipartition condition
between magnetic field and electrons,
implying $B \sim$ $B_{eq}$ $\sim$ 10$^{-5}$ G (e.g., Hardcastle et al. 2002).  
However, the recent case of Pictor A (Migliori et al. 2007) or similar FR\,IIs (e.g.,  Brunetti et al. 2002, Isobe et al. 2005, C05) 
suggest that this requirement is not very tight indicating that for a lobe with $z$ between 0.5 and 1, as
the majority of FRII lobes are detected in X-rays, $B/B_{CMB}$ is of the order of unity (see Figure 1a).

The mean free path, until an electron collides with a magnetic inhomogeneity in a
turbulent plasma, must be smaller than the source size and it has been estimated to be
of the order of $\sim$ 1 kpc (e.g., Lacombe 1977).

Consequently,  substituting $l$ $\sim$ 10$^{-2}$\,$R_{lobe}$ $\simeq$ 10$^{21}$ cm into Equation (3),
and assuming $B/B_{CMB}$ $\sim$ 1, $\gamma_{max}$ is of the order of  10$^5$ - 10$^6$,
as required to interpret the $\gamma$-ray emission of Centaurus A lobes as IC/CMB and
in agreement with the estimate of O'Sullivan et al. (2009).
This is consistent the radio/millimeter observations of radio galaxies (e.g., Hardcastle \& Looney 2008),
that imply the presence of a population of high-energy electrons in their lobes (i.e., $\gamma \sim$ 10$^5$ - 10$^6$).

Our scenario is also in agreement with that proposed by Pacholczyk \& Scott (1976)
for the in situ re-acceleration of extended structures in radio galaxies. 
 
We note that electrons with energies $\gamma$ $\sim$ 10$^7$ - 10$^8$ have synchrotron cooling times $\tau_{syn}$ $\sim$
10$^3$ - 10$^4$ yr and are not expected to be present in radio lobes at the epochs of about few Myrs (which is the typical lobe age).
This is in agreement with the non-detection of lobes at IR and optical frequencies (i.e., $\nu$ $\sim$ 10$^{14}$\,Hz), 
where the synchrotron emission in a magnetic field of about $10^{-5}$ - $10^{-6}$ G of electrons with $\gamma$ $\sim$ 10$^7$ - 10$^8$ should be seen.
%-----------------------------------------------------------------------------------------------------------------------
\begin{figure}[!htp]
\includegraphics[height=7.5cm,width=8.7cm,angle=0]{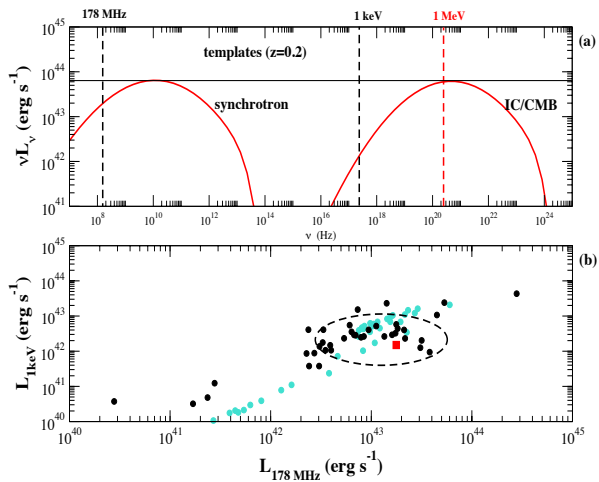}
\caption{a - upper panel) The SED template of a FR\,II lobe adopted to estimate the contribution to the EGB (red line).
b - lower panel) The radio and X-ray luminosities of the lobes in the C05 FR\,II sample (black dots).
The red square indicates the 1 keV and 178 MHz luminosities estimated from our SED template (upper panel). 
The cyan dots show synthetic sources generated
from the model luminosity function describe in $\S$~\ref{sec:egb}.}
\end{figure}
%-----------------------------------------------------------------------------------------------------------------------

\section{The radio lobe spectral energy distribution}
The CMB is a blackbody exhibiting the peak frequency of the SED
at $\nu_{CMB}$ $\sim$ 1.6 $\cdot$ 10$^{11}$ Hz (adopting the Wien law).
Consequently, the expected peak frequency of the 
IC emission in Thomson regime is $\nu_{ic} \sim \nu_{CMB} \, \gamma^2 (1+z)$.
Low energy electrons at $\gamma \sim 10^3$ radiate via IC/CMB
in the X-ray band, while electrons with $\gamma \sim 10^5$ - 10$^6$ 
emit in the MeV - GeV energy range.
The emission arising from highest energy electrons produces the tail of the IC/CMB SED component
and as such its intensity is not expected to be extremely large making
the source undetectable by \fer, unless the radio galaxy lies
at very low redshift.

As previously anticipated this high-energy radiation could contribute to the diffuse EGB. 
To take  this contribution into account, we built a template SED of the lobe
emission from the radio to $\gamma$-rays.
The template has been evaluated assuming a smooth log-normal PED:
$N\left({\gamma}\right)= N_0\, (\gamma/\gamma_p)^{-2-r\log{\left(\gamma/\gamma_p\right)}}$, 
(Massaro et al. 2004). This distribution has been suggested as intrinsic shape of lobe and hotspots (e.g., Katz-Stone \& Rudnick 1994).
The parameters considered to calculate our SED template are reported in Table 1 and it is shown in Figure 2a.
%-----------------------------------------------------------------------------------------------------------------------
\begin{table}
\caption{The parameters of the lobe SED template.}
\begin{tabular}{llcc}
\hline
Parameter  & Symbol & units & template  \\
\hline
\noalign{\smallskip}
redshift                       & z & -- & 0.2 \\
PED curvature          & r & -- & 0.7  \\
PED energy peak     & $\gamma_p$ & -- & 7\,$\cdot$\,10$^3$ \\
PED min energy       & $\gamma_{min}$ & -- & 10$^2$  \\
PED max energy      & $\gamma_{max}$ & -- & $10^6$ \\
PED density              & n & $cm^{-3}$ & 10$^{-7}$ \\
Luminosity ratio        & $L_{syn}/L_{ic}$  & -- & 1 \\
Equivalent Lobe radius & $R_{lobe}$ & Kpc & 45 \\
\noalign{\smallskip}
\hline
\end{tabular}\\
\end{table}
%-----------------------------------------------------------------------------------------------------------------------

We did not assume that the IC/CMB emission is  beamed, since there is no evidence of relativistic bulk motion in lobes.
We chose the value of $R_{lobe}$ to be consistent with the typical value of the C05 sample (see Figure 1(b)), and $\gamma_p$
to have the peak of the synchrotron emission in the radio band at $\sim 10$ GHz. This is justified by  the case of 3C 219 
at $z$ = 0.1744 (Comastri et al. 2003), for which high frequency radio observations are available.
The predicted IR flux of the template at 3.6$\mu$m is $\simeq$ 7 $\times$ 10$^{-18}$ erg\,s$^{-1}$\,cm$^{-2}$, 
in agreement with the non-detected emission at infrared and optical frequencies.

The template is in agreement with the observed radio-to-X-ray luminosities measured
in the lobe sample of C05 (see Figure 2(b) and with the models adopted for the X-ray detected lobes of other FR\,IIs
(e.g., 3C 219, Comastri et al. 2003, 3C~207, Brunetti et al. 2002, 3C~452, 3C~98, 3C~326 Isobe et al. 2002, 2005, 2009, Pictor A, Migliori et al. 2007).

We calculate the SED template with the luminosity ratio between the synchrotron and the IC/CMB component $L_{syn}/L_{ic}$
of the order of unity (Figure 2a). This corresponds to the assumption that the energy losses for the radiative processes considered are 
equivalent, similar condition of $B/B_{CMB}$ $\sim$ 1, in agreement with the observations of FR\,IIs (see Sect. 2). 
In addition, the X-ray spectral index is consistent with the typical slope observed for the few lobes
for which a detailed spectral analysis exists (Hardcastle et al. 2002, C05, Isobe et al. 2002, 2005, 2009).

\section{The contribution to the EGB}
\label{sec:egb}
The estimate of the total diffuse flux arising from the FR\,II radio galaxy population 
can be evaluated adopting the following relation, as reported in Ajello et al. (2008, 2009):

\begin{equation}
F_{EGB}(E_0)=\int dz \int dL \ \Phi (L,z) \frac{dV}{dz} \frac{dN}{dE_0}(E_0(1+z))
\label{eq:diff}
\end{equation}

where $F_{EGB}(E_0)$ is the diffuse flux in units of photons cm$^{-2}$ s$^{-1}$
sr$^{-1}$ MeV$^{-1}$,
$\Phi(L,z)$ is the luminosity function (LF), $\frac{dN}{dE_0}$ is derived from our SED template
of the lobes in the observer's frame at energy $E_0$ and $dV/dz$ is the comoving volume element per unit redshift and 
unit solid angle (Hogg 1999).

The knowledge of the source LF is crucial to estimate correctly this contribution. 
Indeed, it gives the space density of objects with luminosity $L$ at redshift $z$. 
We adopt the FR\,II radio galaxy LF at 15\,GHz (Cara \& Lister 2008).

The SED template used in Equation (4)  is renormalized such that at a given redshift $z$ the luminosity 
in the native radio band of the LF is $L$. Moreover,
the IC component is multiplied by an additional factor $(1+z)^4/(1+0.2)^4$
to account for the fact that the energy density of the CMB scales like $\propto (1+z)^4$. 
%-----------------------------------------------------------------------------------------------------------------------
\begin{figure}[!htp]
\includegraphics[height=6.cm,width=8.5cm,angle=0]{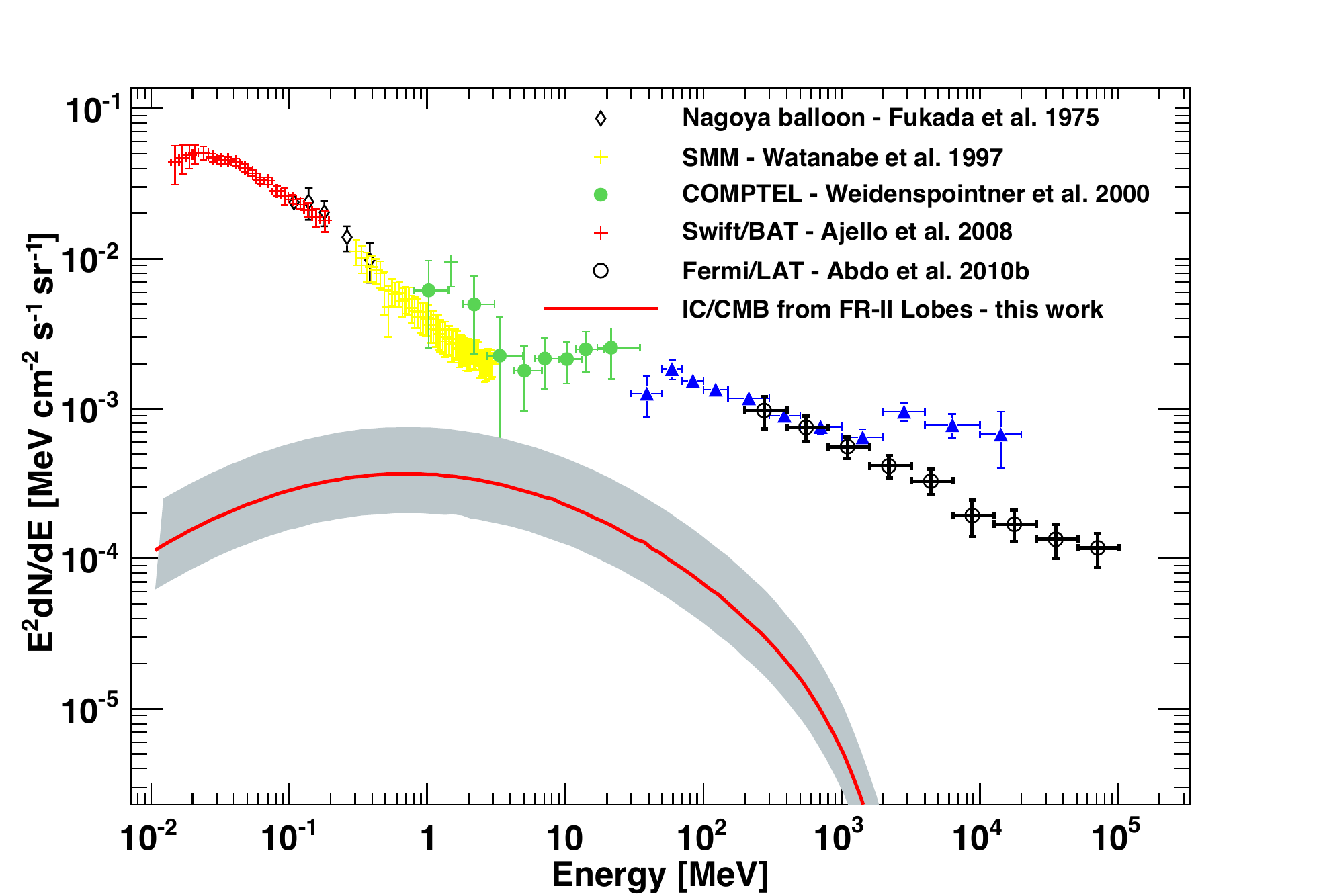}
\caption{The keV to GeV diffuse extragalactic background. 
The red line shows our estimate of the contribution due to the IC/CMB arising from extended lobes of  FR\,II radio galaxies.
The gray area represents the uncertainty in the estimate of the 
contribution to the EGB and arises mainly from
the uncertainties related to the SED template and the estimate of the K constant.
}
\end{figure}
%-----------------------------------------------------------------------------------------------------------------------

The luminosity of the LF of Cara \& Lister (2008) refers
to the luminosity of the FR\,II core while our SED template applies
to the lobe emission. To convert the core luminosity into 
a proxy for the lobe luminosity at 15\,GHz we have used the 
available observations of FR\,II (C05). We have found
that rescaling the core luminosity by a factor $K \approx 0.03$
(and re-normalizing the SED template as described above) produces SEDs
that reproduce well the available data on FR\,IIs for the range of source
luminosities and redshifts reported in C05.
We randomly  extracted sources from our model
and selected only those with flux at 1\,keV $\geq$ 10$^{-15}$\,erg cm$^{-2}$ s$^{-1}$ and $z \leq 1$.
As shown in Figure (2) our LF model is able to reproduce successfully the range of radio and X-ray luminosities
of the C05 sample (see Figure 2(b)).

Moreover, we performed an additional test integrating the LF
coupled to our SED model over luminosity and redshift and we counted
how many objects would be detectable in an all-sky survey
 above a flux of 10$^{-15}$\,erg cm$^{-2}$
s$^{-1}$ in the X-ray band. The number of detectable sources is $\sim$100 
and they display a typical
(average) X-ray luminosity of $5\times10^{42}$\,erg cm$^{-2}$ s$^{-1}$.
These numbers appear to be consistent with the \chn~and \xmm~ observations (e.g., C05).
As an example, adopting a value for $K$ of 10$^{-1}$ or 10$^{-3}$ would produce
$\sim$4000 or $\sim$1 observable sources in the entire sky. Both these scenarios
seem very unlikely given the observations described above. 
We estimated the uncertainties on $K$ adopting the following approach.
To set a lower limit on K, we required that the number of detectable lobes all-sky
(estimated through Equation (4))  at 1\,keV, with flux $\geq$ 10$^{-15}$\,erg cm$^{-2}$ s$^{-1}$ 
matches (within 1\,$\sigma$) the number of sources in the C05 sample.
This condition is satisfied for $K \approx 0.016$. Then, to set
the upper limit, we request that the density of sources observable at X-rays 
(same criteria as above) matches the FR\,II density at 178\,MHz
in the complete sample of Mullin et al. (2008). This condition is satisfied for $K \approx 0.06$.

Fig.~3 shows the contribution to the diffuse EGB emission arising from the 
extended structures of FR\,II lobes and its uncertainty. It is apparent that lobes give a significant,
$\sim$10\,\% contribution to the EGB in the MeV energy range.

\section{Summary and Conclusions}
In this letter, we adopted a statistical acceleration scenario based on turbulent magnetic fields
to show that balancing the particle energy gain with the radiative losses,
the expected Lorentz factor is of the order of $\gamma$ $\sim$ 10$^5$ - 10$^6$
for electrons in the FR\.II lobes, in agreement with the radio/millimeter 
observations of lobes in radio galaxies (e.g., Hardcastle \& Looney 2008).
This has been estimated assuming that the acceleration length scale is comparable to the 
mean free path in the lobe not far from the equipartition condition, supported by the recent
X-ray analyses of C05.

The above scenario is able to justify the presence of high-energy electrons responsible for the
$\gamma$-ray emission arising from the radio galaxy lobes, via IC/CMB, 
as the \fer~observations of the nearby Centaurus A (Abdo et al. 2010a).

Considering the IC/CMB scattering as the main radiative process for the emission in the MeV energy range,
we estimated the contribution of the FR\,IIs to the diffuse EGB.
We found that the peak of this diffuse emission lies at $\sim$ 1~MeV
and the radiation arising from lobes could contribute $\sim$ 10\% of the EGB close to this energy.

So far, the lobes in radio galaxies were only observable at radio frequencies and at high energies in the X-rays. 
This implies that the shape of the emitting PED is uncertain. Consequently, the estimates of the lobe magnetic field
and the possibility of identifying which is the most relevant radiative process, synchrotron or IC emission
are not well constrained. 
However, current models adopted to describe the SED of lobes in FR\,IIs assume the presence of
high-energy electrons up to $\gamma \sim$ 10$^5$ - 10$^6$ (e.g., C05).

To test this hypothesis, very deep infrared and/or hard X-ray observations, with sufficient spatial resolution are necessary.
If \fer~will detect $\gamma$-ray emission arising from nearby FR\,II radio galaxies,
these observations will be crucial to confirm the presence of high-energy electrons in radio lobes
and to constrain their SED.

\vspace{0.2cm}
We thank the anonymous referee for the constructive suggestions that have improved and strengthened the manuscript.
We are grateful to M. Murgia, D. Harris and A. Cavaliere for fruitful suggestions
and to J. Finke, L. Stawarz and T. Cheung for their comments. 
F.~Massaro acknowledges the Foundation BLANCEFLOR Boncompagni-Ludovisi, n'ee Bildt 
for the grant awarded him in 2010.
The work at SAO is supported by the NASA grant NNX10AD50G.
M.~Ajello acknowledges support by NASA grant NNH09ZDA001N.

\end{document}